\documentstyle[epsfig,longtable,graphics]{aipproc}

\begin{document}
\def\today{\ifcase\month\orJanuary\or February\or 
March\or April\or May\or June\orJuly\or August\or 
September\or October\or November\or 
December\fi\space\number\day, \number\year}

\def\pp{\parshape 2 0truecm 15truecm .5truecm 
14.5truecm}
\def\ref #1;#2;#3;#4{\par\pp #1, {\it #2}, {\bf #3}, 
#4}
\def\book #1;#2;#3{\par\pp #1, {\it #2}, #3}
\def\rep #1;#2;#3{\par\pp #1, #2, #3}
\def\undertext#1{$\underline{\smash{\hbox{#1}}}$}
\def\simlt{\lower.5ex\hbox{$\; \buildrel < \over \sim \;$}}
\def\simgt{\lower.5ex\hbox{$\; \buildrel > \over \sim \;$}}
\def\simpropto{\lower.2ex\hbox{$\; \buildrel \propto \over \sim \;$}}
\def\frac #1#2{{#1\over #2}}
\def\deg{\ifmmode^\circ\;\else$^\circ\;$\fi}
\def\solar{\ifmmode_{\mathord\odot}\else$_{\mathord\odot}\;$\fi}
\def\arcmin{\ifmmode^\prime\;\else$^\prime\;$\fi}
\def\arcsec{\ifmmode^{\prime\prime}\;\else$^{\prime\prime}\;$\fi}
\title{ Simulating Galaxy Evolution}

\author{Joseph Silk and  Rychard Bouwens}
\address{Departments of Astronomy and Physics, and Center for Particle
Astrophysics, University of California, Berkeley, CA 94720}

\maketitle

\begin{abstract}
The forwards approach to galaxy formation and evolution is extremely
powerful but leaves several questions unanswered. Foremost among these
is the origin of disks. A backwards approach is able to provide a more
realistic treatment of star formation and feedback and provides a
practical guide to eventually complement galaxy formation \textit{ab
initio}.

\end{abstract}

\section*{Introduction}

Understanding how galaxies formed is the key to unraveling the
mysteries of the high redshift universe.  To interpret the deepest
images of distant galaxies one has to simulate galaxy evolution.  The
prescription for such a simulation seems straightforward.  Take a gas
cloud, massive enough to be self-gravitating, and add a simple
prescription for star formation based on the local free-fall time
scale.  In practice, this approach has yielded star formation
histories that appear to match observations.

However the predictive power of this approach is limited.
The reason is the following.  One has to assume a
prescription for star formation.  Reasonable guesses can be
made, but one has no guarantee that these are valid.  There
is no way of evaluating the uncertainty in the adopted
ansatz for forming stars. This is true for primordial clouds, and
equally valid for current star formation.
Of course the star formation prescription, once selected, has
parameters that can be adjusted, often with little freedom
when confronted with the observational data.  This approach
has been  applied to the early universe, commencing with density
fluctuations that grow by hierarchical clustering of cold
dark matter. 

One can try to assess the uncertainties by comparing snapshots of the
universe at different redshifts.  If one matches the data, one can
deduce that one has a working model of galaxy formation, but one
cannot expect this to be a useful guide to extreme situations that are
not included in the simple algorithm.  These might include, for
example, the role of active galactic nuclei in primordial and current
epoch star formation.  I conclude that it is useful to consider an
alternative to ``{\it ab initio}" galaxy formation.  In this talk I
will describe such an approach that is based on nearby examples of
star formation in a global context, that one attempts to run backwards
in time.  Clearly, forwards and backwards evolution are complementary
descriptions of the same fundamental issues that describe galaxy
formation.

\section*{Galaxy Evolution From Primordial Fluctuations}

Inflationary cosmology prescribes the initial spectrum of
density fluctuations.  The horizon scale at matter-radiation
equality imprints a scale on the relic fluctuation spectrum:
at $L\gg L_{eq}\simeq 12(\Omega\,h^2)^{-1}$ Mpc,
$\delta\rho/\rho\propto M^{-1/2 -n/6}$ and $n\approx 1$
whereas at $L\ll L_{eq}$, $n_{eff}$ approaches $-3$,
reflecting scale invariance for fluctuations that entered
the horizon during radiation domination.  On galaxy scales,
$n_{eff}\approx -2$.  This leads to a hierarchical formation
sequence of structure.  Larger and larger structures merge
together.  Numerical simulations show that some substructure survives.

This is potentially a problem for understanding why galactic
disks remain thin if the surrounding dark halos contain even
a percent of their mass in massive substructures,
characteristic in mass, say, of dwarf galaxies.  Dwarf-disk interactions
would overheat the disk \cite{tot}.
This can be partially rectified by gas infall, which certainly helps
renew thin disks. The discovery of high velocity hydrogen clouds at
the periphery of the halo lends some support to the availability of a
gas reservoir today \cite{bli}.

The properties of dark halos are accounted for by hierarchical
clustering.  The abundance, mass function, density profile and
rotation curve for a typical galaxy halo all agree with empirical
estimates.  The clustering of galaxies is described by the galaxy
correlation function, and simulations of clustering provide a fit over
several decades of scale.  One accounts for the mass function of
galaxy clusters and its evolution with redshift \cite{eke,bah} by
setting $\Omega\approx 0.3\, (\pm 1)$.  Interpretation of massive
halos as rare peaks accounts for the observed clustering of Lyman
break galaxies at $z \sim 3$ \cite{gia}.

The properties of the intergalactic medium agree with predictions of
the hierarchical model.  One has to adopt a metagalactic ionizing
radiation field.  This is taken from the observed quasar luminosity
function. The gas distribution from the simulations is exposed to the
ionizing radiation field, and the effects of the peculiar velocity
field are found to play an important role in reproducing Lyman alpha
cloud absorption profiles. One can explain \cite{kat} the distribution
of observed column densities ranging from damped Lyman alpha clouds
with HI column densities in excess of $10^{21}$ cm$^{-2}$ down to the
Lyman alpha forest below $10^{14}$ cm$^{-2}$.  The gas overdensities
range from $\delta\rho/\rho$ of order several hundred for damped
clouds to unity for the forest.  The structural properties of the
Lyman alpha clouds are simply understood.  There is some controversy
however over the nature of the relatively rare damped clouds.  These
have been argued to be rotating protodisks \cite{wol}. However the
observed spread of velocities is not simply a thin disk, but can
either be interpreted as a thick disk or as a more incoherent,
quasi-spherical halo containing many smaller clouds \cite{hae}.

More problematic for disk theory is the failure of simulations to
reproduce the sizes of galactic disks.  Angular momentum conservation
of a uniformly collapsing and dissipating cloud of baryons within a
dark halo suggests that the disk size is $\lambda R_i$, where $R_i$ is
the halo virialization radius and $\lambda$ is the critical
dimensionless angular momentum.  One has $\lambda\approx 0.06$ and
$R_i$ is typically about 100 kpc.  This argument would actually give
the correct disk size.  However the clumpy nature of the halo is found
to drive efficient angular momentum transfer via dynamical friction.
The disk size found in simulations is a factor of five or more smaller
than observed disk scale lengths \cite{ste}.  Evidently feedback from
star formation is conspiring to limit the collapse of the gas.

The galaxy luminosity function also represents a challenge for
theoretical models, which more naturally specify the galaxy mass
function.  There are two difficulties.  At the low mass end, the
predicted slope of the mass function ($dN/dm\simpropto m^{-2}$) is a
poor fit to the power-law tail of the galaxy luminosity function, the
slope of which depends on galaxy color selection and varies between
$dN/dL\simpropto L^{-3/2}$ in the blue and $dN/dL\simpropto L^{-1}$ in
the red.  One corrects this problem by introducing inefficient star
formation in low mass potential wells.  The fraction of gas forming
stars is assumed to be $(\sigma/\sigma_{cr})^{\alpha}$, with
$\alpha\approx 2$, where $\sigma_{cr} \approx 75\,{\rm km\ s}^{-1}$
denotes the transition velocity dispersion, below which retention of
interstellar gas energised by supernova-driven winds becomes
suppressed.  This assumes that supernovae are effective at disrupting
the interstellar gas in the shallow potential wells characteristic of
dwarf galaxies \cite{dek}.  However the efficiency may only be high at
masses below $\sim 10^6{\rm M}\solar$, according to a recent analysis
of starbursts in dwarf galaxies \cite{mac}.  This would only flatten
the luminosity function at very low luminosities if one counts all
gas-retaining galaxies. One difference between dwarfs ($\simlt
10^8{\rm M}\solar$) and giants is that the supernova ejecta are
expelled, so that the residual gas, if retained, is very
metal-poor. This might be sufficient to reduce the efficiency of star
formation sufficiently so as to produce a population of low surface
brightness dwarfs.

At high luminosities the challenge is to explain why the nonlinear
clustering mass present is $\sim 10^{14}\,{\rm M}\solar$ whereas the
value of L$_*$, above which the number of galaxies decreases
exponentially, is $10^{10}h^{-2}\,\,{\rm L}\solar$.  The corresponding
dark halo mass is around $10^{12}\,{\rm M}\solar$.  Evidently some
physical effect is intervening to limit the luminosity of a galaxy,
which does not track the mass of the dark potential wells.  The
generally accepted resolution is that baryonic cooling is a necessary
condition for star formation to occur in a primordial contracting
cloud.  If the density is too low for gas cooling, the intergalactic
gas remains hot and diffuse.  Efficient star formation must occur
within a dynamical time-scale.  This is certainly how monolithic
formation of an elliptical must have occurred.  In this case, the
condition that the gas cooling time be less than a collapse time sets
a maximum value on cooled galaxy baryon mass of about $10^{12}\,{\rm
M}\solar$.

However gas continues to accrete and cool.  The total mass of cooled
gas does not provide a distinctive cut-off in the mass function of
baryons\cite{tho}.  One has to vary the efficiency of star formation,
reducing it on time-scales longer than a dynamical time, in order to
account for L$_*\,$.  One can appeal to cluster formation to heat up
the intergalactic gas, thereby removing the reservoir of cold gas
which would potentially be accreted.  This would lead one to expect
that cluster ellipticals have a relatively homogeneous distribution of
formation times, peaked at the epoch of cluster formation.  One has to
assume a hot gas environment for field ellipticals, associated with
galaxy groups, to restrict cold gas infall.  However since clustering
in the field develops more recently than for rich clusters one expects
field ellipticals, on the other hand, to display a much broader range
of ages, and reveal, in some cases, signs of recent or current infall.
Indications of this effect can be seen in the enhanced scatter in the
fundamental plane for field ellipticals relative to cluster
ellipticals \cite{for}.

For disk galaxies, the comparison of mass and luminosity via the
predicted mass function challenges interpretations of the Tully-Fisher
correlation between luminosity and maximum rotation velocity
\cite{sten}.  The observed dispersion of fifteen percent in inferred
distance \cite{wil} may be compared with the dispersion between mass
and halo circular velocity in the CDM hierarchy, which is of order 100
percent.  Implementation of a prescription for star formation can
reduce the dispersion between cooled baryon mass, and hence
luminosity, and disk rotational velocity to the observed range by, for
example, allowing stars to preferentially form in the more massive
disks where the baryons are self-gravitating and dense enough to
suppress supernova-driven winds.  However there is a price: the
Tully-Fisher normalization yields the disk mass-to-luminosity ratio,
and the CDM hierarchy inevitably favors a high value relative to the
observed value of $M/L\sim 10\,h$ for the baryon-dominated regions of
disks.

\section*{Galaxy Evolution Via Reverse Engineering}

A complementary approach to galaxy evolution allows one to circumvent
some of these difficulties, although at the risk of introducing other
complications.  One commences with nearby galaxies, develops a model
for star formation, and evolves the galaxies backwards in time.
Actual images or idealized models of nearby galaxies are used as the
starting point.  Suppose one first ignores dynamical evolution.  Star
formation in disks can be described by an expression of the form
$${\rm SFR}=\epsilon\,\mu_{\rm gas}\,\Omega(r)\,f(Q)\,.$$ Here
$\mu_{\rm gas}$ is the surface density of atomic and molecular gas,
$\Omega(r)$ is the rotation rate, $Q$ is the Toomre parameter
(approximately given for a self-gravitating disk of gas by
$\frac{\kappa\sigma}{\pi G \mu_{\rm gas}}$, where $\kappa$ is the
epicyclic frequency and $\sigma$ is the gas velocity dispersion) that
guarantees gravitational instability to axisymmetric perturbations if
$Q<1$, and $\epsilon$ is an efficiency parameter.  One needs to
generalize the dependence on $Q$ to allow for non-axisymmetric
instabilities, such as density waves which are responsible for the
growth of molecular clouds and for the gravitational contribution of
the stellar component.  In general, however, one expects there to be a
threshold for local instabilities when the surface density drops below
a critical value, for typical disks amounting to about $\mu_{\rm
gas}\approx 10\,{\rm M}\solar{\rm pc}^{-2}$.  This empirical
expression fits global star formation rates in disks remarkably well
\cite{ken}, and $\epsilon$ may be interpreted as the fraction of gas
converted into stars per dynamical time.  Infall is one remaining
ingredient that needs to be added.

For individual disks, this model has been exploited to demonstrate
that disks form inside out, that disk surface brightness increases by
almost a magnitude\cite{cay} to $z\sim 1$, and to account for the
chemical evolution of old disk stars and of the interstellar medium at
high redshift\cite{pra}.  The model has considerable potential for
predicting how galaxies appear to evolve in deep images obtained of
the distant universe.  In fact, one study \cite{boub} has already
demonstrated that such a scaling in galaxy size is necessary to
reconcile faint galaxies sizes with galaxies at low redshift, this
study carefully considering changes in the pixelisation, the PSF, and
the surface brightness relative to the noise.  Of course, a careful
consideration of many of the same effects is important for testing
models against the observations.  One has to add a disk formation
epoch, chosen from an analytical prescription for hierarchical CDM
cosmology and some evolution in number density.  The latter is
required to crudely account for merging and is necessary to reproduce
the observed deep galaxy counts.  Ellipticals and spheroids must also
be incorporated into the model.  While these systems do not dominate
the number counts, which at faint magnitudes are dominated by disks
and their irregular precursors, they are important in the cumulative
star formation history of the universe.  Approximately half of the
mass in stars is in the spheroidal component, and hence mostly in E's
and S0's.  This is the approximate assessment for the local luminosity
function (and is due to the fact that while $\sim 30$ percent of
galaxies are E's and S0's, the associated $M/L$ is about twice as
large as for typical spirals).  One also reaches an independent
verification of this from the cosmic far infrared background.  This
recently discovered diffuse flux at 100 -- 300 $\mu$m amounts to
$\lambda i_{\nu} \approx 20\,{\rm nw/m}^2/{\rm s}^2$, comparable to
the diffuse optical light flux when integrated over the HDF and near
infrared.  Modelling of disk galaxies incorporating dust can reproduce
the optical background but only about fifty percent of the FIR
background is explained by optically visible systems.  The remainder
is presumed to be due to dusty ellipticals.  Of course if these
systems form stars at an early epoch $z_E$ relative to spirals at
$z_S$, then the inferred mass in stars (for the same initial mass
function) in dust enshrouded spheroids is equal to $[(1+z_E)/(1+z_S)]$
times the contribution from disks.  This comparison suggests that $z_E
\sim z_S$ though in principle one could have $z_E\gg z_S$.

One might worry that the FIR background could be due to AGN.
However modelling of the x-ray background effectively
constrains the AGN contribution to diffuse hard photons.
Compton self absorption of the x-rays, required to obtain a
spectral fit of the XRB, limits the possible contribution to
the diffuse FIR background by dust-shrouded, x-ray-emitting
AGN to be almost ten percent of the observed background.
Direct observations by SCUBA find ultraluminous galaxies at
$z=1\,$--$\,3$.  Perhaps ten percent of these may be AGN-powered
according to the previous argument, and this is consistent
with direct spectroscopic signatures.

\subsection*{Disk Parameterisation}

There are two major uncertainties in the modelling of the disk star
formation rate: infall and efficiency.  One can constrain the role of
infall by three independent methods that respectively appeal to
chemical evolution, disk dynamics, and to the evolution of disk sizes.
The best studied is chemical evolution.  Infall of metal-poor gas into
the early gas is required to account for the paucity of metal poor G
dwarfs.  The sharp decline in supersolar metallicities of disk stars
means that recent metal-poor infall is greatly reduced relative to
infall in the first 5 Gyr.  Infall of gas-rich clumps is predicted in
the CDM model, but these interactions must avoid overheating the disk.
Less than 4\% of the disk can have fallen in over the past 5 Gyr
according to one study \cite{tot}.  However recent calculations
suggest that infalling satellites preferentially tilt rather than heat
the disk \cite{car}.  The implications for high redshift galaxies is
that disks are small at $z>1$.  Without infall, disks would not be
sufficiently small, according to one recent analysis, to account for
the decrease in faint galaxy angular diameter.

One can only decompose disks from bulges to $z\simlt 1$,
using HST data.  Evolution of disk sizes to this redshift is
quite model-dependent.  Disk size varying as
$(1+z)^{\alpha}$, with $\alpha \approx 2$, fits the available
data.  However selection biases need to be modelled more
carefully.  One selects earlier type galaxies at high
redshift than at low redshift because of surface brightness
dimming, and this complicates comparisons.

\subsection*{Disk Physics}

The essence of disk formation lies in inefficiency.
Galaxies retain a sufficient gas reservoir so as to still be
vigorously forming stars at the present epoch.  The star
formation rate increases dramatically with cosmic epoch,
possibly peaking near $z\sim 2$.  Hence gas infall drops off
dramatically.  This also is implicit in models of galactic
chemical evolution, where infall of metal-poor gas over the
first five or so Gyr helps account for the metal
distribution of old disk stars.  The inefficiency of star
formation must be due, not to the availability of a gas
supply but rather arises from being controlled by disk
physics.

Feedback of energy and momentum from star formation and
death necessarily play an important role.  One needs to
include such physics to understand disk sizes.  One could
simultaneously account for gas longevity.  Angular momentum
transfer is central to such a model.  A general class of theories
which can successfully reproduce disk profiles is based on
contracting viscous self-gravitating disks.  The viscosity
arises from cloud-cloud collisions, the cold disk being
gravitationally unstable to cloud formation.  The disk forms
as angular momentum is transferred on a viscosity timescale.
Since cloud collisions and mergers are assumed to drive star
formation, one naturally relates the star formation and
viscosity time scales.  An exponential surface density
profile is naturally generated \cite{lin}.

\subsection*{Bulge Evolution}

Bulges are expected to be prominent in observations of high redshift
galaxies, both because of disk evolution and the high bulge surface
brightness.  Yet the sequence of bulge formation is poorly understood,
and this makes it difficult to formulate and test {\it ab initio}
predictions of disk evolution.  Consider the following alternatives.
Bulges form before disks, either monolithically or in major
(i.e. comparable, or at least mass ratio 1:10) mergers.  Bulges form
simultaneously with disks via satellite mergers.  Bulges form after
disks, via secular instability of disks, and bar formation followed by
dissolution as gas inflow drives bulge formation \cite {nor}.  Any of
these scenarios are possible.  Two, or even all three, may be
operative.  For example, secular evolution can form small bulges but
not the massive objects of early type galaxies. Observational evidence
that bulge and disk scale lengths are correlated favors a secular
evolution origin of bulges for late-type spirals \cite{courteau}.  The
ubiquity of bars, which are efficient at torquing accreting gas and
driving the gas inwards to form a central bulge, also suggests that
secular evolution must have played a significant role in bulge
formation.  Conversely, massive bulges are most likely formed by
mergers.  Satellite infall of gas-rich dwarf galaxies is expected to
be a common occurrence in hierarchical models and provides a natural
mechanism for simultaneously forming the bulges, as the dense stellar
cores sink into the center of the galaxy by dynamical friction, and
feeding disk growth with gas infall.  There are hints of monolithic
bulge formation from observations of many compact Lyman break
galaxies, which have high star formation rates.

One can try to address this confusing range of bulge formation
possibilities by examining the properties of disk galaxies at $z
\simlt 1$, where component separation into bulge and disk is possible
at HST resolution.  Late-forming bulges are inevitably bluer and
smaller than early-forming bulges, at a given redshift.  Figure 1
shows a comparison of the model predictions with available data.  HST
images are shown, in a comparison with the HDF.  Analyses of similar
images \cite{boucs} show that only with larger samples at $z\sim 1$
could one be able to distinguish between alternative models of bulge
formation.

\section*{Looking to the Future}

It will be possible in the not too distant future to greatly refine
the observational constraints relevant to galaxy evolution.  In Figure
2 we show HST Advanced Camera (2000) and NGST (2007) simulations of
the same 85'' x 85'' field using the secular evolution scheme for
bulge formation.  The Advanced Camera simulations consider a 150,000-s
integration, utilise a pixel size of 0.05 arcsec, and probe the $gri$
optical bands to $i_{AB} \sim 30.3$, whereas the NGST simulations
consider a similar 150,000-s integration, utilise a pixel size of
0.029 arcsec, and probe the 1,3,5-$\mu$m wavelength bands to $m_{1\mu
m,AB}\sim 31.6$.  For comparison, we also show WFPC2 (pixel size is
0.1 arcsec, probes the $I_{F814W}$, $V_{F606W}$, and $B_{F450W}$ bands
to a limiting magnitude $I_{F814W,AB} \sim 29$) and NICMOS (pixel size
is 0.2 arcsec, probes the $J_{F110W}$ and $H_{F160W}$ infrared bands
to $H_{F160W,AB} \sim 28.3$) simulations.  Since the fiducial secular
model for bulge formation breaks down at high redshift, we have
included a variation of the Pozzetti, Bruzual, \& Zamorani \cite{poz}
luminosity evolution model at these redshifts.  The simulations
include both $K$ and evolutionary corrections, cosmological angular
size relations and volume elements ($\Omega = 0.15$, $h=0.5$),
appropriate pixelisation, PSFs, and noise (see \cite{boub} for a
discussion).

Obviously, one of the principal advantages of the Advanced Camera and
NGST over WFPC2 and NICMOS are their increases in limiting magnitude,
angular resolution, and field of view.  Regarding the differing
limiting magnitudes, using the Advanced Camera for similar length
exposures to those shown here, one could probe to unobscured star
formation rates $\sim0.5 M_{\odot}/yr$ at $z\sim5$ whereas with WFPC2,
the limiting rate is only $\sim2 M_{\odot}/yr$.  For higher redshift
observations, such as are only possible with NICMOS or NGST, NGST
promises to push the sensitivity on unresolved star formation from its
current value $\sim20 M_{\odot}/yr$ at $z\sim10$ obtainable with
NICMOS exposures down to $\sim1 M_{\odot}/yr$.

\begin{figure}
\resizebox{15cm}{!}{\includegraphics*[120,204][500,605]{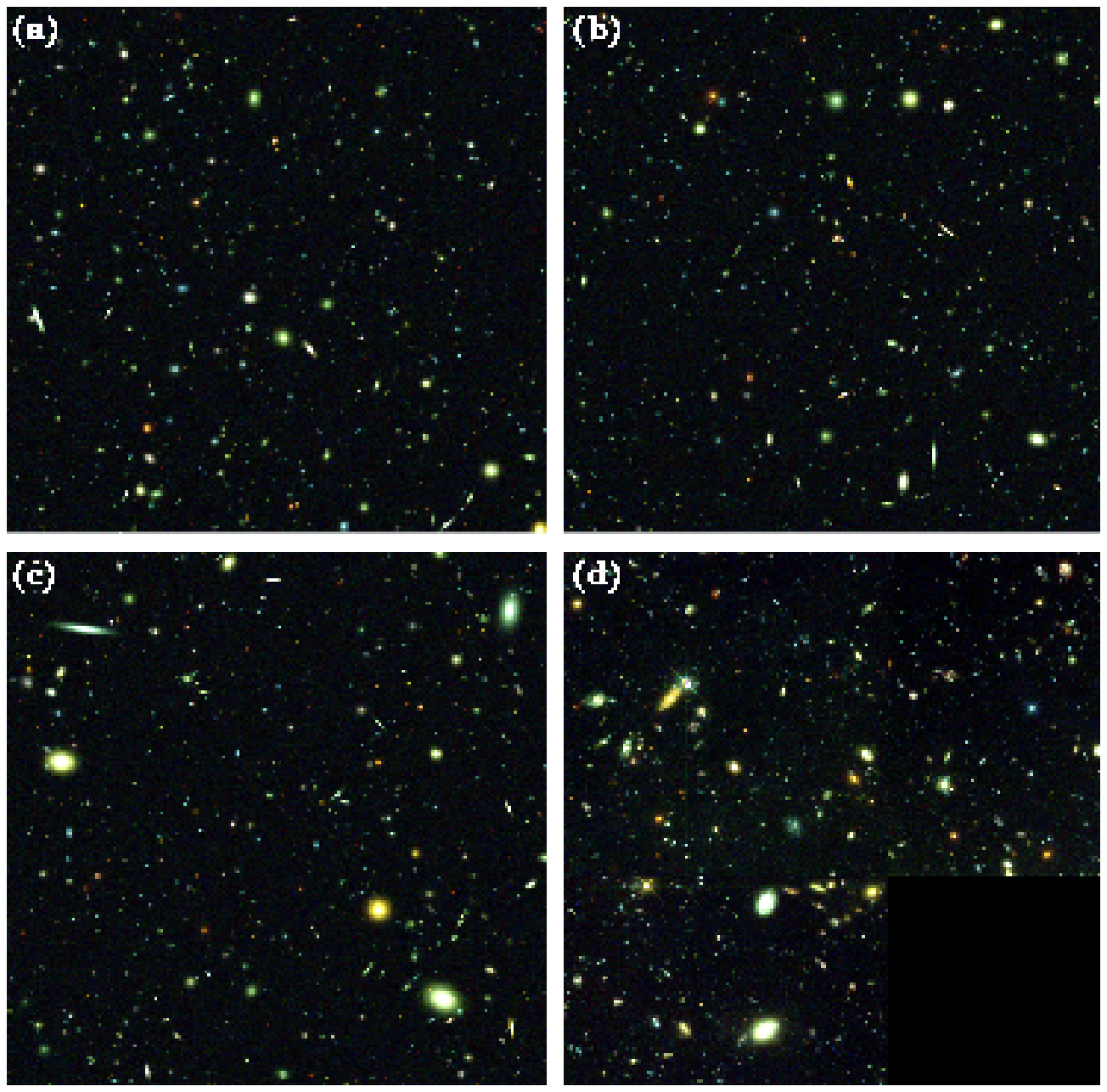}}
\caption{Comparison of simulated BVI images of a 2'' x 2'' patch of
the HDF with the observed images (panel d).  Panel (a) illustrates our
secular evolution model for bulges, panel (b) illustrates our
simultaneous formation model, and panel (c) illustrates our early
bulge formation model.  Calculations are performed using a
galaxy-evolution software package written by one of the authors.}
\end{figure}

\newpage

\begin{figure}
\resizebox{15cm}{!}{\includegraphics*[120,204][500,605]{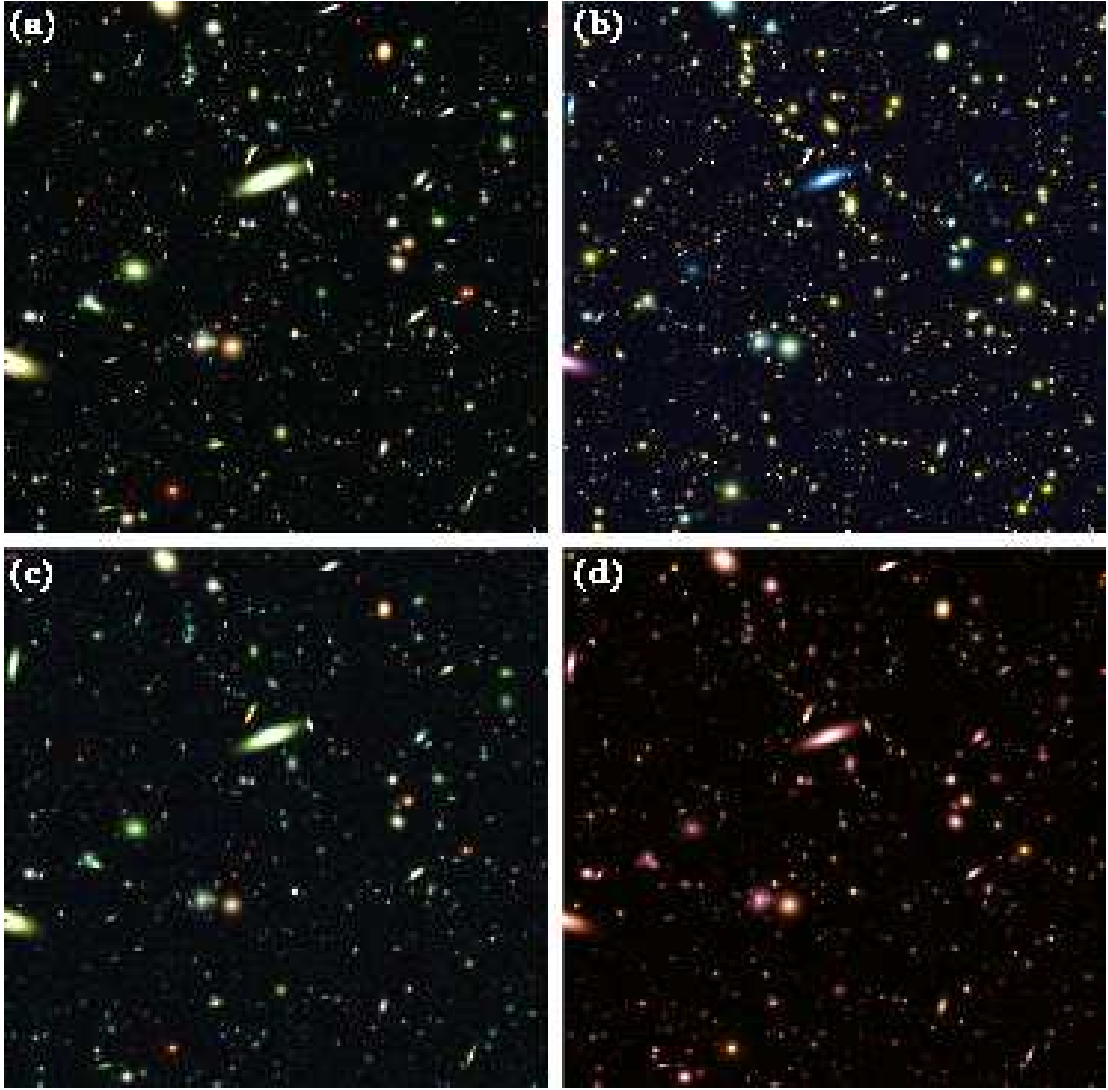}}
\caption{Simulated images of a 85'' x 85'' field using a secular
evolution model for disks to $z=1$ and the Pozzetti, Bruzual, \&
Zamorani luminosity evolution model for $z>1$.  Shown are 30-orbit
$gri$ exposures for the HST Advanced Camera (a), 150000-s 1,3,5-$\mu$m
exposures for NGST (b), 30-orbit $BVI$ exposures for HST WFPC2 (c),
and 30-orbit $JH$ exposures for the HST NIC3 camera (d).  Calculations
are performed using a galaxy-evolution software package written by one
of the authors.}
\end{figure}

\end{document}